\documentclass[conference]{IEEEtran}
\IEEEoverridecommandlockouts
\usepackage[left=1.62cm,right=1.62cm,top=1.9cm]{geometry}
\usepackage{cite}
\usepackage{amsmath,amssymb,amsfonts}
\usepackage{algorithmic}
\usepackage{graphicx}
\usepackage{textcomp}
\usepackage{amsmath}

\usepackage{amsmath}
\usepackage{mathtools}
\usepackage{verbatim}
\usepackage{xcolor}

\usepackage[T1]{fontenc}
\usepackage[utf8]{inputenc}
\usepackage{authblk}

\def\BibTeX{{\rm B\kern-.05em{\sc i\kern-.025em b}\kern-.08em
    T\kern-.1667em\lower.7ex\hbox{E}\kern-.125emX}}
\begin{document}

\title{Active Reconfigurable Intelligent
Surface-Aided Terahertz Wireless Communications\\

{%\footnotesize \textsuperscript{*}Note: Sub-titles are not captured in Xplore and should not be used
} \thanks{This work was partly supported by “Regional Innovation Strategy (RIS)” through the National Research Foundation of Korea (NRF) funded by the Ministry of Education (MOE) (2021RIS-004), Basic Science Research Program through the NRF funded by the MOE (NRF-2022R1I1A1A01071807, 2021R1I1A3041887).}
}

\author[*]{Waqas Khalid}
\author[$\dagger$]{Heejung Yu}
\author[$\ddagger$]{Yazdan Ahmad Qadri}
\affil[*]{Institute of Industrial Technology, Korea University, Sejong, Korea; waqas283@\{korea.ac.kr, gmail.com\}}
\affil[$\dagger$]{Dept. of Elec. \& Inform. Eng., Korea University, Sejong, Korea; heejungyu@korea.ac.kr}
\affil[$\ddagger$]{Department of Computer Science, Yeungnam University, Gyeongsan-si, Korea; yazdan@yu.ac.kr}
\renewcommand\Authands{ and }

%\author{\IEEEauthorblockN{Waqas Khalid}
%\IEEEauthorblockA{\textit{Institute of %Industrial Technology} \\
%\textit{Korea University}\\
%Sejong 30019, South Korea \\
%waqas283@korea.ac.kr; waqas283@gmail.com}
%\and

%\IEEEauthorblockN{M. Atif Ur Rehman}
%\IEEEauthorblockA{\textit{Dept. of %Computing \& Mathematics}\\
%Manchester Metropolitan University\\
%Manchester M1 5GD, UK\\
%m.atif.ur.rehman@mmu.ac.uk}
%\and
%\IEEEauthorblockN{Heejung Yu}
%\IEEEauthorblockA{\textit{Dept. of Elec. \& %Inform. Eng.} \\
%\textit{Korea University}\\
%Sejong 30019, South Korea \\
%heejungyu@korea.ac.kr}

%}

\maketitle

\begin{abstract}
\small Terahertz (THz) communication is expected to be a key technology for future sixth-generation (6G) wireless networks. Furthermore, reconfigurable intelligent surfaces (RIS) have been proposed to modify the wireless propagation environment and enhance system performance. Given the sensitivity to blockages and limited coverage range, RIS is particularly promising for THz communications. Active RIS can overcome the multiplicative fading effect in RIS-aided communications. In this paper, we explore active RIS-assisted THz communications. We formulate the ergodic rate, considering factors associated with active RIS, including active noise and signal amplification, and THz signals, including molecular absorption and beam misalignment.
\end{abstract}

\begin{IEEEkeywords}

THz, Active RIS, signal amplification, molecular absorption, beam misalignment.
\end{IEEEkeywords}

\section{Introduction}

Terahertz (THz) communication leverages frequencies in the range of 0.1 to 10 THz, offering ultra-wide bandwidth and high-speed data transmission. Nonetheless, THz signals encounter severe energy absorption and increased path loss, primarily due to molecular absorption caused by their short wavelengths. Moreover, the directional nature of THz antennas poses another significant challenge, namely, beam misalignment \cite{b0}.

Reconfigurable intelligent surfaces (RIS) have emerged as an enabling technology in 6G networks due to their ability to actively reshape wireless environments \cite{b1}. Through the manipulation of low-cost passive reflecting elements, RIS introduces additional cascaded channels alongside the direct link, effectively enhancing communication performance. Compared to active relays, RIS offers significant advantages including lower energy consumption, flexible deployment, minimal noise, and cost-effectiveness. In RIS-aided communications, the path loss of the cascaded channel is the product of the path losses from the BS-RIS and RIS-user links. This multiplicative fading phenomenon reduces the capacity gain provided by the RISs and constrains their practical applications. To address this challenge, active RIS has been proposed, which involves integrating a power amplifier within each RIS element. Active RIS amplifies the reflected signal, thereby mitigating the multiplicative path loss effect and bolstering communication performance \cite{b2}.

\subsection{Motivation}

The consideration of molecular absorption and beam misalignment in THz communications, along with signal amplification and active noise for active RIS, is essential for accurately predicting signal behavior and optimizing system design parameters. However, further investigation for analytical modeling incorporating these key factors in active-RIS-aided THz communications is required.

\subsection{Contribution}

In this paper, we explore active RIS-assisted THz communications, formulating the ergodic rate while considering key factors such as active noise, signal amplification, path loss, molecular absorption, and beam misalignment.

\begin{figure}[t]
\centering
\includegraphics[width=2.8in,height=2in]{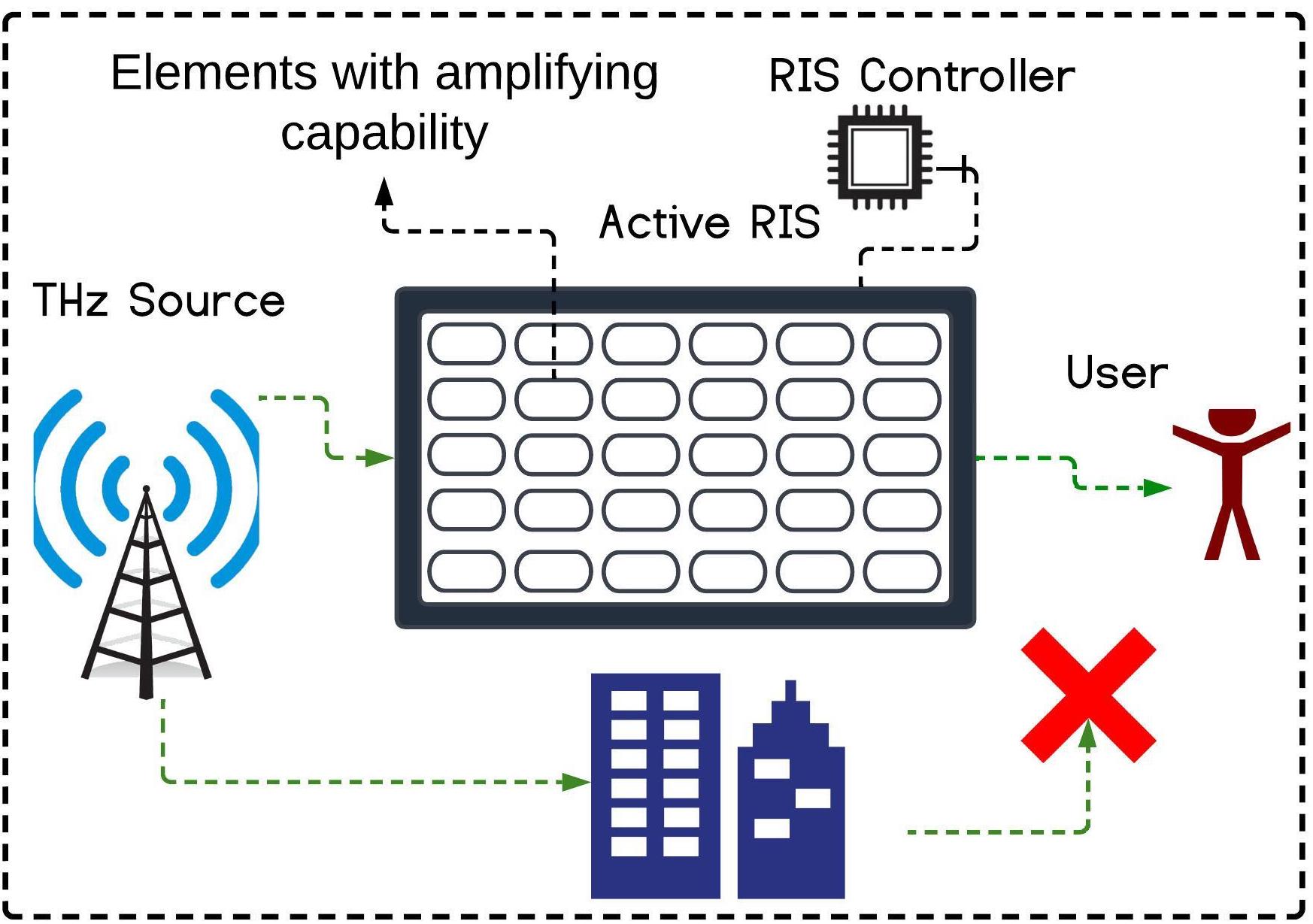}
\caption {System model of the active RIS-assisted THz communications.}
\label{diagram1}
\end{figure}

\section{System Model}

We examine a scenario involving a THz communication system aided by a RIS, assuming the direct link between the base station (BS) and the user $U$ is obstructed. The BS employs an RIS with $M$ elements to establish communication with $U$. Given the substantial path loss in THz communications due to molecular absorption and beam misalignment, the signal power reflected multiple times by the RIS is assumed negligible. The fading channel coefficients between the BS and RIS, and between the RIS and $U$, denoted as $f_m$ and $g_m$ respectively, follow complex Gaussian distributions \cite{b3}.

\subsection{Path Gain Coefficient}

The path gain coefficient can be evaluated as $h_L=h_P h_{A}$, where $h_p$ is the propagation gain and $h_{A}$ is the molecular absorption gain. We can express $h_P$ as follow:

\begin{align}\label{equ:1}
h_P=\frac{c^2\sqrt{G_a G_b}}{\left(4\pi f\right)^2 d_a d_b}
\end{align}
where $G_a$ and $G_b$ denote the antenna gains, $c$ is the speed of light, $f$ represents the frequency, and $d_a$ and $d_b$ indicate the BS-RIS distance and RIS-$U$ distance, respectively.

Furthermore, $h_{A}$ can be calculated as,

\begin{align}\label{equ:1}
h_{A}=\exp{\left(-\frac{\kappa(f)\left(d_a+d_b\right)}{2}\right)}
\end{align}
where $\kappa(f)$ denotes the molecular absorption coefficient \cite{b0}.

\subsection{Beam Misalignment Coefficient}

Let $h_{M}$ represent the beam misalignment with its probability density functions (PDFs) is written as \cite{b0},

\begin{equation} \label{eq5}
f_{h_{M}} (x) = \zeta \phi^{-\zeta} x^{\zeta-1}, \quad 0 \leq x \leq \phi,
\end{equation}
where $\phi = (\operatorname{erf}(s))^2$ is the power captured by $U$ under perfect alignment conditions. $\zeta = \frac{v^2}{4 \sigma^2}$; $v$ is the beam-width and $\sigma^2$ is the variance of the beam misalignment. Moreover, $\operatorname{erf}(\cdot)$ is the error function. $s = \sqrt{\frac{\pi}{2}} \frac{r}{u}$; $r$ is the radius of the $U$'s effective area and $u$ is the BS's beam footprint.

\section{Performance Analysis}

The received signal at $U$ is given by,

\begin{align} \label{eq4}
y=\sqrt{P_s}h_{L}h_{M}\sum^{M}_{m=1}f_mg_m\beta e{^{j\theta_m}}x+\beta n_r+n_u
\end{align}
where $\theta_m$ is the induced phase for the $m$th element, $\beta$ is the amplification factor for each element (which may exceed one for the active RIS), $x$ represents the transmitted signal from the BS, $P_s$ is the transmission power, and $n_r$ and $n_u$ are the thermal noises at the active RIS and $U$, respectively.

The ergodic capacity of $U$ can be expressed as, 

\begin{align} \label{eq5}
C_u=\frac{1}{\ln2}\int_0^\infty\frac{1-F_\gamma(s)}{1+s}ds
\end{align}
where $\gamma$ is the SNR at $U$, which can be expressed as,

\begin{align} \label{eq6}
\gamma= \rho_sh_{L}^2\beta^2|h_{M}| ^2\;\left(\sum^{M}_{m=1}|f_m| |g_m|\right)^2
\end{align}
where $\rho_s=\frac{P_s}{\beta^2\sigma^2_{r}+\sigma^2_{u}}$.

To determine Eq. \ref{eq5}, we utilize the cumulative distribution function (CDF) of $\chi = \left(\sum_{m=1}^M |f_m| |g_m|\right)^2$, denoted as $F_\chi(s)$. Using the moment matching technique, $\chi$ can be approximated by a Gamma distribution with shape parameter $k$, and scale parameter $\omega$. Thus, the CDF $F_\chi(s)$ can be written as,

\begin{align} \label{eq7}
F_\chi(s) = 1-\frac{\gamma\left(k, \frac{s}{\omega}\right)}{\Gamma(k)}
\end{align}
where $k=\frac{\left(\mathbb{E}\left\{\chi\right\}\right)^2}{\mathbb{V}\left\{\chi\right\}} $, and $\omega=\frac{\mathbb{V}\left\{\chi\right\}}{\mathbb{E}\left\{\chi\right\}}$. 

Assuming $|f_m|$ and $|g_m|$ follow a Rayleigh distribution, the statistical values can be calculated as: $\mathbb{E}\left\{\sum_{m=1}^M |f_m| |g_m|\right\} = M \frac{\pi}{4}$, $\mathbb{V}\left\{\sum_{m=1}^M |f_m| |g_m|\right\} = M \left(1 - \frac{\pi^2}{16}\right)$, $\mathbb{E}\left\{ \left(\sum_{m=1}^M |f_m| |g_m|\right)^2\right\}=\left( M \frac{\pi}{4}\right)^2+M \left(1 - \frac{\pi^2}{16}\right)$ and $\mathbb{E}\left\{ \left(\sum_{m=1}^M |f_m| |g_m|\right)^4\right\}=\left( M \frac{\pi}{4}\right)^4+\left( M \frac{\pi}{4}\right)^2M \left(1 - \frac{\pi^2}{16}\right)+\left(M \left(1 - \frac{\pi^2}{16}\right)\right)^2$.

Finally, the conditional CDF of $F_\gamma$ can be expressed as,

\begin{align} \label{eq8}
F_\gamma(s \vert  x) = 1-\frac{1}{\Gamma(k)}\gamma\left(k, \frac{s}{\rho_sh_{L}^2 x^2 \omega}\right)
\end{align}

The unconditional CDF of $F_\gamma$ can be written as,

\begin{align} \label{eq9}
F_\gamma (s) =1- \int_{0}^{\phi} \frac{\zeta  x^{\zeta-1}}{\phi^{\zeta} \Gamma(k)}\gamma\left(k, \frac{s}{\rho_sh_{L}^2 x^2 \omega}\right)dx
\end{align}

These equations can then solve for the ergodic capacity of $U$ as detailed in Eq. \ref{eq5}.

\begin{comment}
\section{Numerical Results}
To verify the accuracy of the expressions, we present numerical results. Unless specified otherwise, the parameters are set as: $G_a$=$G_b$=30 dBi, $d_a$=$d_b$= 15 m, $f$=0.3 THz, $M$=100, $\beta$=2, $s$=0.3, $\zeta$=.6, $P_s$=30 dBm, and $\sigma^2_{r}$=$\sigma^2_{u}$= 0.01.  Active RISs have shown great potential for enhancing THz communications by addressing the challenges associated with THz signals, including limited coverage range and severe losses, as well as the multiplicative fading effect in traditional RIS communications. 
\end{comment}

\section{Conclusion}
Active RISs have shown great potential for enhancing THz communications by addressing the challenges associated with THz signals, including limited coverage range and severe losses, as well as the multiplicative fading effect in traditional RIS communications. In our study, we investigate the use of active RISs for THz communications and develop an analytical framework for the ergodic rate, taking into account aspects of active RISs, such as active noise and signal amplification. Additionally, we consider factors related to THz signals, such as molecular absorption and beam misalignment.

%
%\section*{Acknowledgment}

%Conceptualization, W.Khalid, H. Yu,; writing—original-draft preparation, writing—review and editing. All
%authors have read and agreed to the published version of the manuscript.

%\section*{References}

\end{document}